\documentclass[10pt]{iopart}

\usepackage{iopams}  
\usepackage{graphicx,epsfig}
\usepackage{amssymb}
\newcommand{\ba}{\begin{eqnarray}}
\newcommand{\ea}{\end{eqnarray}}

\begin{document}

\title{Electromagnetic transitions in the algebraic cluster model}

\author{R. Bijker}
\address{Instituto de Ciencias Nucleares, 
Universidad Nacional Aut\'onoma de M\'exico, 
A.P. 70-543, 04510 M\'exico, D.F., M\'exico}
\ead{bijker@nucleares.unam.mx}
\author{O. A. D{\'{\i}}az-Caballero}
\address{Facultad de Ciencias, 
Universidad Nacional Aut\'onoma de M\'exico, 
A.P. 70-543, 04510 M\'exico, D.F., M\'exico}

\begin{abstract}
We study electromagnetic transition rates in the framework of the algebraic cluster model. 
The concept of shape-phase transitions is used to propose a mechanism that allows to have 
interband and intraband quadrupole transitions of comparable strength, as observed in $^{12}$C.  
\end{abstract}


\vspace{2pc}

\ioptwocol

\section{Introduction} 

Recently there has been a lot of renewed interest in $\alpha$ clustering in light nuclei 
\cite{FreerFynbo}. Especially, the measurement of new excited states in $^{12}$C has motivated 
many theoretical studies to understand the structure of $\alpha$-cluster nuclei, like the  
semi-microscopic algebraic cluster model \cite{Cseh}, the antisymmetrized molecular dynamics 
\cite{AMD}, fermionic molecular dynamics \cite{FMD}, BEC-like cluster model \cite{BEC}, 
{\it ab initio} no-core shell model \cite{NCSM}, lattice effective field theory \cite{EFT}, 
no-core symplectic model \cite{Draayer}, and the algebraic cluster model (ACM) \cite{ACM,O16}. 
Recent reviews on the structure of $^{12}$C can be found in Ref.~\cite{FreerFynbo,Tohsaki} and
on $\alpha$-particle clusters in \cite{Schuck}.  

The algebraic cluster model for three-body clusters was introduced in 2000 in an application 
to $^{12}$C \cite{ACM}. In the ACM, the cluster states in $^{12}$C are interpreted in terms of 
rotations and vibrations of a (equilateral) triangular configuration of $\alpha$ particles. 
As a consequence of the geometric symmetry positive and negative parity states merge into a 
single ground-state rotational band characterized by the sequence $L^P=0^+$, $2^+$, $3^-$, 
$4^{\pm}$, $5^-$, all of which have been observed by now, most recently the $4^-$ state in 
2007 \cite{Fre07} and the $5^-$ states in 2014 \cite{C12}. 
In the ACM, the Hoyle state is interpreted as the bandhead of a breathing vibration of the 
triangular configuration, on top of which an entire rotational band is built. Since the 
underlying geometric symmetry is the same as that of the ground state, the rotational 
sequence is expected to be the same as that of the ground-state band. In recent years, 
evidence has been found for the $2^+$ state of the Hoyle band in three different experiments 
\cite{Itoh,Freer,Gai}, as well as for the $4^+$ Hoyle state \cite{Fre11}. 

In 2014, the ACM was extended to four-body clusters. An application to the cluster states 
in $^{16}$O suggested that these can be interpreted in terms of rotations and vibrations of 
tetrahedral configuration of $\alpha$ particles. 
The triangular configuration in $^{12}$C and the tetrahedral configuration in $^{16}$O 
implied by the observed rotational sequences, were confirmed by a study of $B(EL)$ electric 
transitions along the ground-state band \cite{ACM,O16}

The main discrepancy concerns the electromagnetic decay of the Hoyle state in $^{12}$C. 
In the ACM, the quadrupole transition from the Hoyle state to the first excited $2^+$ state 
is an interband transition which is suppressed with respect to intraband transitions 
whereas experimentally they are found to be comparable. 

The aim of this contribution is to present a possible solution to this problem. As a first 
step we analyze the spectral properties of a simplified version of the ACM namely for two-body 
clusters. It will be shown that a Hamiltonian which mixes the spherical and the deformed phases 
can have intra- and interband quadrupole transitions of comparable strength 
while still retaining, to a good approximation, the rotational structure of the bands. 
 
\section{The algebraic cluster model}

For the case of three identical clusters, as is the case for $^{12}$C as a cluster of three $\alpha$ 
particles, the ACM Hamiltonian has a spherical phase and a deformed phase in which the clusters are 
located at the vertices of an equilateral triangle. In Ref.~\cite{ACM} the latter scenario was used to 
study the spectral properties of $^{12}$C. The Hoyle state was interpreted as a breathing vibration of 
the triangular configuration. As a consequence, the quadrupole transition from the Hoyle state to the 
ground-state band corresponds to an interband transition which is suppressed with respect to transitions 
within the same band, see Table~\ref{ACMC12}. 

\begin{table}
\centering
\caption{Comparison of ratios of excitation energies and $B(E2)$ values in $^{12}$C 
\cite{ajzenberg} and the calculated values in the three-body ACM \cite{ACM}.}
\vspace{10pt}
\label{ACMC12}
\begin{tabular}{ccc}
\hline
\noalign{\smallskip}
& Exp & ACM \\
\noalign{\smallskip}
\hline
\noalign{\smallskip}
$\frac{E(4_1^+)}{E(2_1^+)}$ & $3.17$ & $3.33$ \\
\noalign{\smallskip}
$\frac{B(E2;0_2^+ \rightarrow 2_1^+)}{B(E2;2_1^+ \rightarrow 0_1^+)}$ 
& $1.72 \pm 0.25$ & $0.15$ \\
\noalign{\smallskip}
\hline
\end{tabular}
\end{table}

Whereas the spherical phase corresponds to a dynamical symmetry of the ACM in which all nuclear 
properties of interest can be obtained in closed analytic form, the triangular configuration does 
not. In order to investigate the transition between a spherical and a deformed phase, we analyze 
the spectral properties of a simplified version of the ACM, namely for two-body clusters, which 
has the advantage that both the spherical and the deformed phases correspond to a dynamical symmetry. 
This makes it possible to analyze selection rules and derive closed expressions for quadrupole 
transitions which can be used a benchmarks for the two phases.  

\subsection{Two-body ACM}

Originally, the two-body ACM was introduced in applications to diatomic molecules as the vibron 
model \cite{FI}. The vibron model is an algebraic model to describe the relative motion of two 
clusters. The building blocks are four bosons, the three components of a vector boson with $L^P=1^-$ 
and a scalar boson with $L^P=0^+$, denoted by
\ba 
b^{\dagger}_{m} ~, \; s^{\dagger} ~, 
\ea
with $m=\pm 1$, $0$. The group structure is $U(4)$. The model space is characterized by the symmetric 
irreducible representation $[N]$ of $U(4)$ where $N$ represents the total number of bosons. 

The most general one- and two-body ACM Hamiltonian that conserves the total number of bosons, 
angular momentum and parity, is given by 
\ba
H &=& \epsilon_{0} \, s^{\dagger} s - \epsilon_{1} \, b^{\dagger} \cdot \tilde{b} 
+ u_0 \, s^{\dagger} s^{\dagger} s s 
\nonumber\\
&& - u_1 \, s^{\dagger} b^{\dagger} \cdot \tilde{b} s
+ v_0 \left[ b^{\dagger} \cdot b^{\dagger} s s
+ s^{\dagger} s^{\dagger} \tilde{b} \cdot \tilde{b} \right] 
\nonumber\\
&& + \sum_{L=0,2} a_{L} \, ( b^{\dagger} \times b^{\dagger} )^{(L)} \cdot 
( \tilde{b} \times \tilde{b} )^{(L)} ~, 
\label{HS2}
\ea
with $\tilde{b}_m=(-1)^{1-m}b_{-m}$. 
For two identical clusters, the condition of invariance under the interchange of the two clusters 
is equivalent to parity conservation. Hence, positive parity state are symmetric and negative parity 
states antisymmetric. Since we do not consider internal excitations of the clusters, the two-body 
wave functions arise from the relative motion only, and have to be symmetric. Therefore, the allowed 
states are the ones with positive parity. 

The ACM Hamiltonian of Eq.~(\ref{HS2}) has two special solutions corresponding to dynamical 
symmetries called the $U(3)$ and $SO(4)$ limit, which can be interpreted as the three-dimensional 
harmonic oscillator and the deformed oscillator, respectively \cite{onno,RB}. In the following, 
we study the spectral properties of these two cases.

\subsection{Harmonic oscillator}

The first dynamical symmetry corresponds to the group chain 
\ba
\left| \begin{array}{ccccccccc}
U(4) &\supset& U(3) &\supset& SO(3) \\
N &,& n &,& L \end{array} \right> ~.
\ea
The basis states are classified by the the total number of bosons $N$, the number of oscillator quanta 
$n=0,1,\ldots,N$, and the angular momentum $L=n,n-2,\ldots,1$ or $0$ for $n$ odd or even. The parity 
of the levels is given by $P=(-1)^{n}$. For two identical clusters, the allowed states have $n$ even, 
and therefore also $L$ even.  

We consider the one-body Hamiltonian 
\ba
H_1 = \epsilon \sum_{m} b_{m}^{\dagger} b_m ~, 
\label{ham1}
\ea
with eigenvalues  
\ba
E_1 = \epsilon \, n ~. 
\label{e1}
\ea
The corresponding energy spectrum is that of a three-dimensional harmonic oscillator.  

In the ACM, $E2$ transitions are described by the quadrupole operator 
\numparts
\ba
Q^{(2)} &=& q_2 \, \left[ D \times D \right]^{(2)} ~, \\
\hat D_{m} &=& (b^{\dagger} s - s^{\dagger} \tilde{b})^{(1)}_m ~. 
\ea
\endnumparts
In the $U(3)$ limit, the selection rule of the quadrupole operator is given by $\Delta n=0$, $\pm 2$. 
The ratio of $B(E2)$ values can be derived in closed analytic form as
\numparts
\ba
\frac{B(E2;0_2^+ \rightarrow 2_1^+)}{B(E2;2_1^+ \rightarrow 0_1^+)} &=& \frac{10(2N-3)^2}{3N(N-1)} 
\nonumber\\ 
&=& \frac{40}{3} \left[ 1-\frac{2}{N}+{\cal O}(1/N^2) \right] ~, \\ 
\frac{B(E2;4_1^+ \rightarrow 2_1^+)}{B(E2;2_1^+ \rightarrow 0_1^+)} &=& \frac{6(N-2)(N-3)}{N(N-1)} 
\nonumber\\ 
&=& 6 \left[ 1-\frac{4}{N}+{\cal O}(1/N^2) \right] ~.
\ea
\endnumparts

\subsection{Deformed oscillator}

The second dynamical symmetry corresponds to the group chain 
\ba
\left| \begin{array}{ccccccccc}
U(4) &\supset& SO(4) &\supset& SO(3) \\
N &,& \sigma &,& L \end{array} \right> ~. 
\ea
The basis states are classified by the quantum numbers $N$, $\sigma$ and $L$, 
In this case, the energy levels are organized into bands labeled by $\sigma$ with 
$\sigma =N,N-2,\ldots,1$ or $0$ for $N$ odd or even, respectively. For two identical 
clusters, the rotational excitations are denoted by $L$ even with $L=0,2,\ldots,\sigma$. 

In this case, the Hamiltonian is given by 
\ba
H_{2} = \xi \, P^{\dagger} P + \kappa \, \vec{L} \cdot \vec{L} ~,
\label{H2b}
\ea
where $P^{\dagger}$ denotes a generalized pairing operator 
\ba
P^{\dagger} = s^{\dagger} s^{\dagger} - b^{\dagger} \cdot b^{\dagger} ~, 
\label{pair2}
\ea
and $\vec{L}$ is the angular momentum in coordinate space 
\ba
\hat{L}_m = \sqrt{2} \, ( b^{\dagger} \times \tilde{b} )^{(1)}_m ~.
\ea
The energy eigenvalues are given by
\ba
E_2 = \xi (N-\sigma)(N-\sigma+2) + \kappa L(L+1) ~,
\ea
which corresponds to the rotation-vibration spectrum of a three-dimensional deformed oscillator, 
consisting of a series of vibrational excitations labeled by $\sigma$, and rotational states 
labeled by $L$.  

\begin{figure}
\centering
\setlength{\unitlength}{0.7pt} 
\begin{picture}(300,260)(0,0)
\thinlines
\put (  0,  0) {\line(1,0){300}}
\put (  0,260) {\line(1,0){300}}
\put (  0,  0) {\line(0,1){260}}
\put (150,  0) {\line(0,1){260}}
\put (300,  0) {\line(0,1){260}}
\thicklines
\put ( 50, 60) {\line(1,0){20}}
\put ( 50,120) {\line(1,0){20}}
\put ( 50,180) {\line(1,0){20}}
\put (100,120) {\line(1,0){20}}
\put ( 60,120) {\vector(0,-1){60}}
\put ( 60,180) {\vector(0,-1){60}}
\put ( 85,120) {\oval(50,50)[tr]}
\put ( 85,120) {\oval(50,50)[tl]}
%\put ( 60,125) {\vector(0,-1){5}}
\thinlines
\put ( 10, 57) {$n$=0}
\put ( 10,117) {$n$=2}
\put ( 10,177) {$n$=4}
\put ( 75, 57) {$0_1^+$}
\put ( 75,117) {$2_1^+$}
\put ( 75,177) {$4_1^+$}
\put (125,117) {$0_2^+$}
\put (100,220) {$U(3)$}
\thicklines
\put (170, 60) {\line(1,0){20}}
\put (170, 90) {\line(1,0){20}}
\put (170,160) {\line(1,0){20}}
\put (220,140) {\line(1,0){20}}
\put (180, 90) {\vector(0,-1){30}}
\put (180,160) {\vector(0,-1){70}}
\put (230,140) {\vector(-1,-1){50}}
\put (205,117) {\bf +}
\thinlines
\put (170, 30) {$\sigma$=$N$}
\put (195, 57) {$0_1^+$}
\put (195, 87) {$2_1^+$}
\put (195,157) {$4_1^+$}
\put (240,120) {$\sigma$=$N$-2}
\put (245,137) {$0_2^+$}
\put (235,220) {$SO(4)$}
\end{picture}
\caption{Characteristic features of the spectrum in the $U(3)$ limit 
(left) and the $SO(4)$ limit (right).} 
\label{sphdef}
\end{figure}
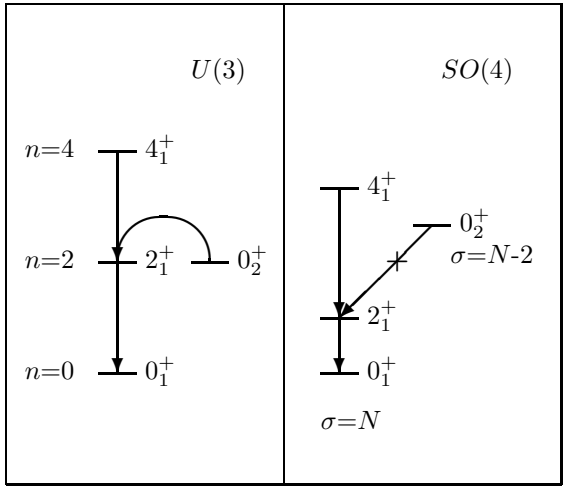

In the $SO(4)$ limit, the selection rule of the quadrupole operator $Q^{(2)}$ is given by 
$\Delta \sigma=0$, {\it i.e.} only intraband transitions are allowed. The ratios of $B(E2)$ 
values can be derived as  
\numparts
\ba
\frac{B(E2;0_2^+ \rightarrow 2_1^+)}{B(E2;2_1^+ \rightarrow 0_1^+)} &=& 0 ~, \\
\frac{B(E2;4_1^+ \rightarrow 2_1^+)}{B(E2;2_1^+ \rightarrow 0_1^+)} 
&=& \frac{10(N-2)(N-3)(N+4)(N+5)}{7N(N-1)(N+2)(N+3)}
\nonumber\\ 
&=& \frac{10}{7} \left[ 1-\frac{20}{N^2}+{\cal O}(1/N^3) \right] ~.
\ea
\endnumparts
In this case, the quadrupole transition $B(E2;0_2^+ \rightarrow 2_1^+)$ corresponds to an interband 
band transition which is forbidden by the $\Delta \sigma=0$ selection rule.
 
The characteristics of the two dynamical symmetries are summarized in Fig.~\ref{sphdef} and 
Table~\ref{BE2}. In the $U(3)$ limit the spectrum is harmonic, and the ratio of the excitation 
energies of the first excited $4^+$ and the first excited $2^+$ states is $E(4_1^+)/E(2_1^+)=2$, 
whereas in the $SO(4)$ limit these two states belong to a rotational band with energy ratio $10/3$. 

The main difference between the two dynamical symmetries concerns the quadrupole transition 
between the first excited $0^+$ and the first excited $2^+$ state. For the harmonic oscillator 
both states belong to the same $n=2$ multiplet and therefore the $B(E2;0_2^+ \rightarrow 2_1^+)$ 
is allowed by the $\Delta n=0$ selection rule. 
For the deformed oscillator, the first excited $2^+$ state belongs to the ground-state rotational band 
with $\sigma=N$, whereas the first excited $0^+$ state corresponds is the bandhead of the breathing 
vibration with $\sigma=N-2$. As a consequence, the quadrupole transition is forbidden by the 
$\Delta \sigma=0$ selection rule. The ratio of $B(E2)$ values 
\ba
\frac{B(E2;0_2^+ \rightarrow 2_1^+)}{B(E2;2_1^+ \rightarrow 0_1^+)} ~,
\ea
changes from $0$ in the deformed oscillator to $40/3$ (to leading order in $N$) for the harmonic 
oscillator, in comparison to the value of $1.72 \pm 0.25$ as measured in $^{12}$C, see 
Table~\ref{ACMC12}. 

\begin{table}
\centering
\caption{Ratios of excitation energies and $B(E2)$ values to leading order in $N$ 
in the $U(3)$ and $SO(4)$ limits of the two-body ACM.}
\vspace{10pt}
\label{BE2}
\begin{tabular}{ccc}
\hline
\noalign{\smallskip}
& $U(3)$ & $SO(4)$ \\
\noalign{\smallskip}
\hline
\noalign{\smallskip}
$\frac{E(4_1^+)}{E(2_1^+)}$ & 2 & $\frac{10}{3}$ \\
\noalign{\smallskip}
$\frac{B(E2;0_2^+ \rightarrow 2_1^+)}{B(E2;2_1^+ \rightarrow 0_1^+)}$ 
& $\frac{40}{3}$ & 0 \\
\noalign{\smallskip}
\hline
\end{tabular}
\end{table}

The situation for $^{12}$C is inbetween the results for the $U(3)$ and $SO(4)$ limits. 
The spectrum is to a good approximation rotational with $E(4_1^+)/E(2_1^+)=3.17$, whereas 
the strenghts of the quadrupole transitions $B(E2;0_2^+ \rightarrow 2_1^+)$ and 
$B(E2;2_1^+ \rightarrow 0_1^+)$ are comparable. 

In the next section we will mix the two scenarios to investigate the transitional 
region between the $U(3)$ and $SO(4)$ limits in a systematic manner, to see whether 
there exists an intermediate Hamiltonian that reproduces the ratios of energies and 
$B(E2)$ values as observed in $^{12}$C.  

\section{Shape-phase transition}
\label{phases}

The transitional region between the harmonic and the deformed oscillator limits can be 
described by the schematic Hamiltonian \cite{onno,RB}
\ba
H = (1-\chi) \sum_{m} b_{m}^{\dagger} b_{m} 
+ \chi \left[ \frac{1}{4(N-1)} \, P^{\dagger} P + \kappa \vec{L} \cdot \vec{L} \right] ~, 
\nonumber\\
\mbox{}
\label{shapes}
\ea
with $0 \leq \chi \leq 1$. For $\chi=0$ it reduces to the harmonic oscillator or $U(3)$ limit 
and for $\chi=1$ to the deformed oscillator or $SO(4)$ limit. Under the assumption that 
three-cluster data might be used in this two-body cluster model, the value of $\kappa$ is chosen 
such as to reproduce for $\chi=1$ the observed energy ratio $E(0_2^+)/E(2_1^+)=1.72$ in $^{12}$C. 

\begin{figure}
\centerline{\epsfig{file=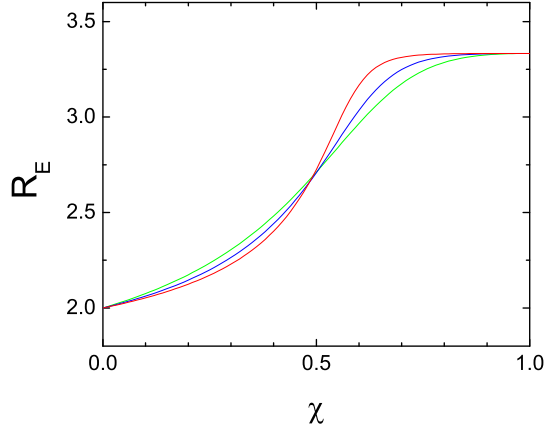,width=\linewidth}}
\caption{The energy ratio $R_E=E(4_1^+)/E(2_1^+)$ as a function of $\chi$ 
for $N=6$ (green), $N=10$ (blue) and $N=20$ (red).}
\label{RE}
\end{figure}

\begin{figure}
\centerline{\epsfig{file=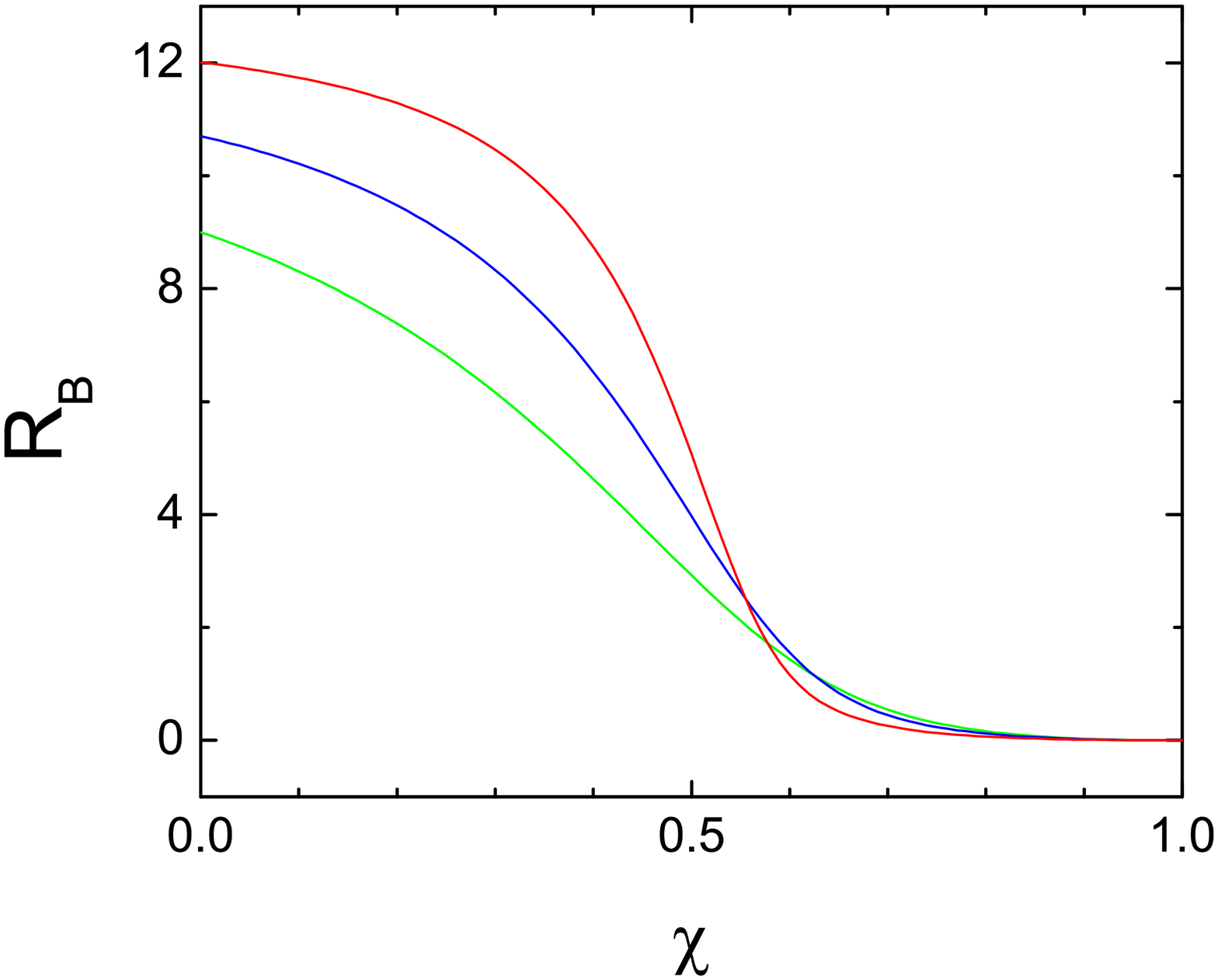,width=\linewidth}}
\caption{As Fig.~\ref{RE}, but for the ratio of $B(E2)$ values 
$R_{B}=B(E2;0_2^+ \rightarrow 2_1^+)/B(E2;2_1^+ \rightarrow 0_1^+)$.}
\label{RB1}
\end{figure}

The schematic Hamiltonian of Eq.~(\ref{shapes}) exhibits a second-order phase 
transition at the critical point $\chi_c=1/2$ \cite{onno,RB}. 
Figures~\ref{RE} and \ref{RB1} show the results for the energy ratio $E(4_1^+)/E(2_1^+)$ 
and the ratio of $B(E2)$ values $B(E2;0_2^+ \rightarrow 2_1^+)/B(E2;2_1^+ \rightarrow 0_1^+)$ 
in the transitional region for three different values of $N$: $N=6$ (green), $N=10$ (blue) and 
$N=20$ (red). The value of $N=10$ was used in a earlier study of $^{12}$C in the  
three-body ACM \cite{ACM}. 

Inspection of Figures~\ref{RE} and \ref{RB1} shows that the conditions for $^{12}$C are met 
approximately for a value of $\chi$ relatively close to the critical value, $\chi \sim 0.6$. 
For example, for $N=10$ bosons (the number used in \cite{ACM}) and $\chi=0.59$ the results 
are $R_E=3.00$ for the energy ratio and $R_B=1.74$ for the ratio of $B(E2)$ values. For this 
value of $\chi$ the spectrum is, to a good approximation, still rotational whereas the 
$B(E2)$ values for intraband and interband quadrupole transitions are comparable.  

\begin{figure}
\centerline{\epsfig{file=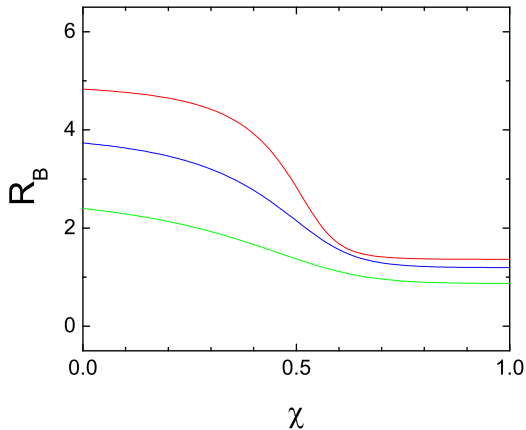,width=\linewidth}}
\caption{As Fig.~\ref{RE}, but for the ratio of $B(E2)$ values 
$R_{B}=B(E2;4_1^+ \rightarrow 2_1^+)/B(E2;2_1^+ \rightarrow 0_1^+)$.}
\label{RB2}
\end{figure}

Finally, in Figure~\ref{RB2} we show the results for a ratio of two quadrupole transitions 
within the ground-state band $B(E2;4_1^+ \rightarrow 2_1^+)/B(E2;2_1^+ \rightarrow 0_1^+)$. 
In the large $N$ limit, this ratio varies between the vibrational value $6$ and the 
rotational value of $10/7$.  

\section{Summary and conclusions}

In this article, we studied the quadrupole transitions in the algebraic cluster model. 
We used a schematic Hamiltonian for two-body clusters which describes the transitional region 
between the spherical and the deformed phases. Particular attention was paid to the behavior of 
the ratio of inter- and intraband transitions. In the deformed limit, interband transitions 
are forbidden, whereas in the spherical limit, the transition between the $0_2^+$ and $2_1^+$ 
states is much larger than the one between the $2_1^+$ and the ground state. 
It was shown that by mixing the two phases with a value of the control parameter $\chi \sim 0.6$ 
which is a bit larger than the critical value $\chi_c=1/2$, the two transitions are of the 
same order while the energy spectrum is still close to rotational. The first results are very 
promising, and may offer a mechanism to explain the observed electromagnetic properties of the 
Hoyle state in $^{12}$C in the framework of the ACM by mixing the deformed triangular phase 
with a spherical term, such that the quadrupole transition between the Hoyle state and the 
$2_1^+$ state is large and comparable to quadrupole transitions inside the ground-state band
\cite{work}. 

\section*{Acknowledgments}

This work was supported in part by research grant IN109017 from PAPIIT-DGAPA, UNAM.

\end{document}